\documentclass[aps,10pt,pra,twocolumn,amsmath,amssymb,superscriptaddress]{revtex4-1}

\usepackage[english]{babel} 
\usepackage[utf8]{inputenc}
\usepackage[T1]{fontenc}

\usepackage{lipsum}

\usepackage{bbm}
\usepackage{verbatim}
\usepackage{graphicx}
\usepackage{times}
\usepackage{epsfig}
\usepackage{bm}
\usepackage{txfonts}
\usepackage{dsfont}
\usepackage{color}

\usepackage{hyperref}
\hypersetup{
colorlinks=true,
linkcolor=blue,         
citecolor=blue,         
filecolor=magenta,      
urlcolor=red
}

\begin{document}

\selectlanguage{english}

\title{
Divergence of predictive model output as indication of phase transitions
}
 
 \author{Frank Sch\"afer }
\affiliation{Department of Physics, University of Basel, Klingelbergstrasse 82, CH-4056 Basel, Switzerland}

\author{Niels L\"orch}
\email{niels.loerch@unibas.ch}
\affiliation{Department of Physics, University of Basel, Klingelbergstrasse 82, CH-4056 Basel, Switzerland}

\date{\today}

\begin{abstract}
We introduce a new method to identify phase boundaries in physical systems. It is based on training a predictive model such as a neural network to infer a physical system's parameters from its state.
The deviation of the inferred parameters from the underlying correct parameters will be most susceptible and diverge maximally in the vicinity of phase boundaries. Therefore, peaks in the divergence of the model's predictions are used as indication of phase transitions. Our method is applicable for phase diagrams of arbitrary parameter dimension
and without prior information about the phases. 
Application to both the two-dimensional Ising model and the dissipative Kuramoto-Hopf model show promising results.
\end{abstract}


\maketitle

\section{Introduction}

Recent developments in machine learning demonstrating the surprising predictive power of deep neural networks \cite{schmidhuber2015, lecun2015, goodfellow2016} have led to great renewed interest in the field and its potential to revolutionize both industry and academic disciplines.
Breakthrough achievements responsible for this optimism include unprecedented accuracy in image classification \cite{krizhevsky2012}, 
super-human skills at strategy games like go and chess \cite{silver2016}, or even creation of art \cite{goodfellow2014}.

This technology is being transferred to physics and has been applied e.g. in high-energy physics to interpret of experimental data at the LHC \cite{kasieczka2017} and in astronomy to recover features in images of galaxies \cite{schawinski2017}.
In particular, in quantum and statistical mechanics,
there has been great interest to integrate machine learning with quantum technology \cite{schuld2015, biamonte2017},  with successful
design and control of experiments \cite{melnikov2018, fosel2018, bukov2018}, optimization of numerical algorithms \cite{huang2017, liu2017} and efficient representation of physical systems  with Boltzmann machines \cite{carleo2017, koch2018}. 

Feed-forward neural networks have been proven highly efficient in labeling different phases of matter \cite{carrasquilla2017}, even without knowing the correct labels beforehand \cite{van2017}, manifesting a kind of unsupervised learning for the discovery of phase transitions \cite{wang2016}. 
Further interesting techniques for this purpose include methods building on features of other predictive models such as support vector machines \cite{liu2018muenchen}
 or on unsupervised clustering algorithms \cite{wang2016, wetzel2017, hu2017, ch2018}, e.g.  principal component analysis (PCA),
 t-SNE or k-means clustering.


The schemes presented in Refs. \cite{van2017, van2018} predict phase transitions by retraining a neural network on a system with tentative phase labels many times, to accept the particular partition where the network achieves highest labeling accuracy as the prediction for the correct separation of phases. Refs. \cite{broecker2017, liu2018} generalize this approach to two-dimensional phase diagrams, where this procedure becomes more costly.
Ref.~\cite{huembeli2018} can find phase boundaries between two different phases, circumventing the cost of network retraining, by leveraging adversarial domain adaption \cite{ganin2016domain} to make use of prior knowledge of correct phase labels in some well-understood area of phase space.

Here, we introduce a new method that can naturally predict arbitrary-dimensional phase diagrams  without any prior knowledge of phase labels.
It is economical in computational resources, as it only requires one training procedure, where any suitable predictive model is taught to infer from the state of a physical system its system parameters.   
The deviation of the inferred parameters from the underlying correct parameters can then be used to predict phase transitions:
 As the model's predictions will tend to be most susceptible to the change of system parameters in the vicinity of phase boundaries, this is where the divergence of predictions will peak and thereby indicate phase transitions.
As the method does not require any prior knowledge of the labels or even number of different phases, our algorithm constitutes an unsupervised learning scheme, while employing a supervised subroutine learning the labeled system parameters.

We will demonstrate the method on two test cases. Firstly, to start with a well-understood system, we apply it to
the Ising model on a two-dimensional lattice, with the potential variation of anisotropic coupling in horizontal and vertical direction.
As the second test case we choose to study an example showing a dissipative
non-equilibrium phase transition. As these are generally harder to describe theoretically, it is especially important to develop new methods for a better understanding.
In particular, we consider the Kuramoto-Hopf model, an effective model that was recently established \cite{marquardt2015}
 to describe the effective dynamics of weakly coupled nonlinear self-oscillators on a two-dimensional lattice, as may be experimentally implemented with optomechanical systems.

The structure of this article is as follows:
In the subsequent section, we introduce our divergence-based learning scheme for uncovering phase transitions and explain its mechanism in theory. We will then review the two physical systems used as test cases and apply the scheme to them. Finally we discuss the potential and limitations of the method and give an outlook for potential future developments.

\section{Rationale for the Learning scheme}

\begin{figure}[t!]
\includegraphics[width=0.5\textwidth]{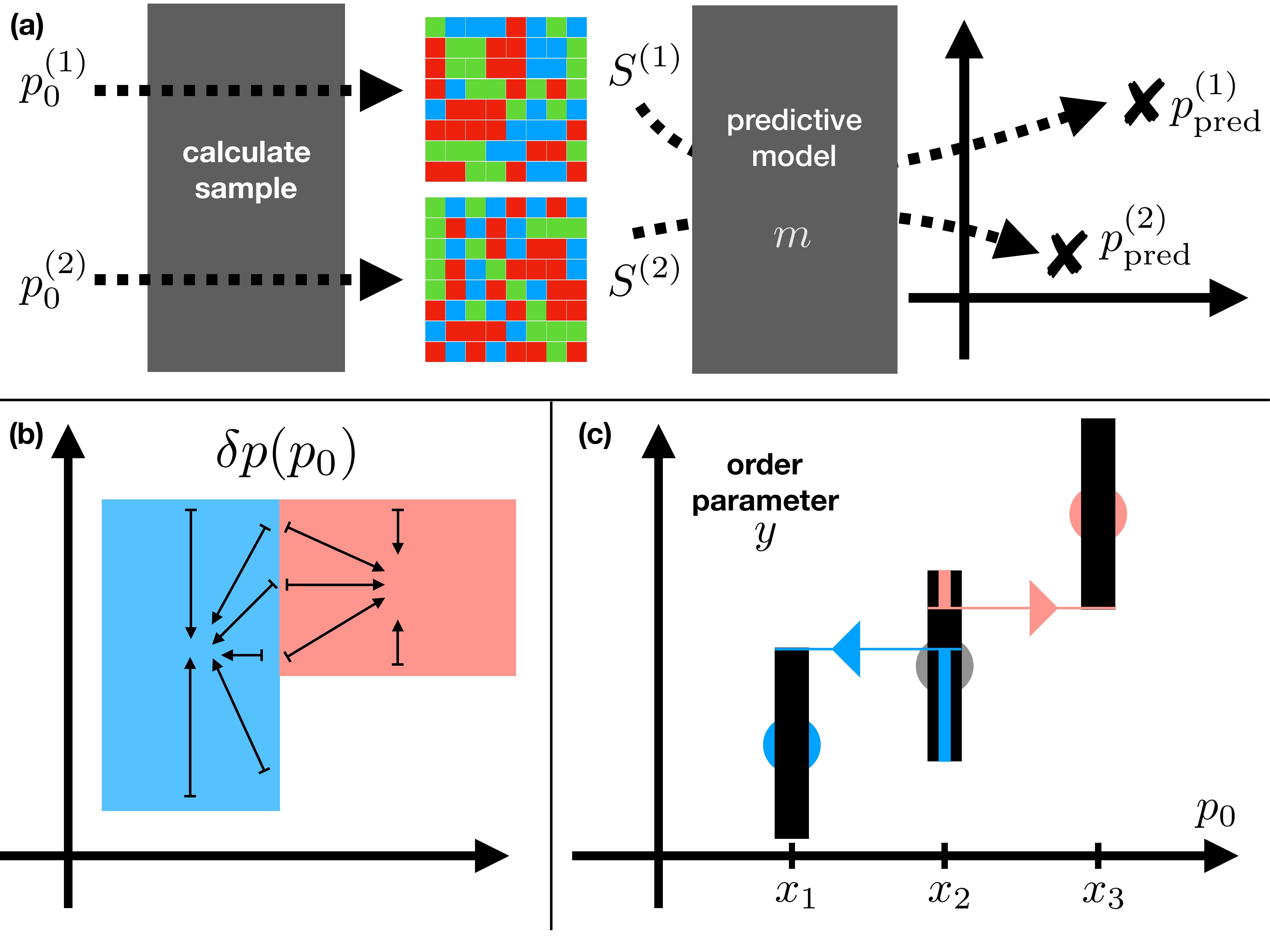}
\caption{Illustration of the divergence-based model to find phase transitions. (a) Parameters $p_0^{(j)}$ are uniformly drawn from parameter space to calculate corresponding samples $S^{(j)}$. A predictive model is trained to recognize the parameters for a given sample, such that the deviation $\delta p^{(j)}$ of the resulting prediction $p_{\mathrm{pred}}^{(j)}$ from the real $p_0^{(j)}$ is minimized on average. If the model can distinguish parameters between different phases better than within a phase,
crossing the border between different phases results in diverging predictions $\delta p (p_0)$, which provides a signature of a phase transition. Panel (b) illustrates this phenomenon by showing the vector field $\delta p (p_0)$ in the presence of two phases (blue and red) for the limit of zero resolution within a phase, where all predictions within a phase will be placed at its center of mass. 
The analytical argument leading to Eq.~\eqref{eq:condition} for the opposite limiting case of high resolution is illustrated in panel (c), where the black bars correspond to the range of possible estimates of the order parameter at a given point, and the colored bars indicate the overlap to the neighboring bars. In this limit, the prediction will deviate
towards the direction of higher overlap.
}
\label{fig:method}
\end{figure}

Our method requires to uniformly sample instances of the state $S$ of a physical system as a function of continuous system parameters $p_0$. For example, $S(p_0)$ could be a spin configuration of the Ising model sampled as a function of temperature and coupling strengths as the parameters.

As illustrated in Fig.~\ref{fig:method}(a), the ansatz of our model is as follows: A predictive model $m: S \to p_{\mathrm{pred}}$, for example a neural network, is trained to produce predictions $p_{\mathrm{pred}}$ of the system parameters that minimize the expectation value of the loss function
\begin{align}
 L=\langle (\delta p)^2 \rangle ,
\label{eq:loss}
\end{align}
where $\delta p=p_{\mathrm{pred}}-p_0$ denotes the difference between the predictions and the correct labels
\footnote{While other choices of $L$ are possible, here we chose the generic Euclidian norm (mean squared error) of the difference.}.

For sufficiently precise models, the divergence 
\begin{align}
\label{eq:div}
\mathrm{div} ({\delta p}) = \sum_n \frac{  \partial {\delta p_n} }{{\partial p_n}}
\end{align}
will then generally have local maxima at the parameter values, where the system state is most susceptible to a change of system parameters. This susceptibility suggests that the system undergoes a phase transition at those parameters. We note that, as $\mathrm{div}({p_0})$ is constant, we could equivalently use the maxima of $\mathrm{div}({ p_{\mathrm{pred}}}) = \mathrm{div}({\delta p}) +\mathrm{div}({p_0})$.

The working of this scheme can be intuitively understood by the following considerations. The best strategy to minimize the mean squared errors within a patch of parameter space that looks indistinguishable to the predictive model, is to place the prediction at the center of mass of that patch. Even if such a patch can be resolved by the model to some extent, there will still be a bias at the edges of the patch towards the center, as long as the model's precision and confidence are not absolute.
Therefore, two neighboring patches that are very well distinguishable from each other will lead to diverging predictions in the vicinity of the border separating these patches and therefore the maximum of the divergence described above. 
This argument is illustrated in Fig.~\ref{fig:method}(b) for the limiting case of low resolution, where the predictive model is incapable of resolving the parameters within a phase, but is able to discriminate between states of different phases. 

In the opposite limit of a predictive model with extremely high resolution, we quantify the expected value of $\delta p$ after making a few idealized assumptions. While this quantification helps to better understand our scheme, one should also keep in mind that it describes a particular idealized situation.

We assume that the phase transitions of the considered system can be described with an order parameter $y(p)$, which in the following will be a real number for simplicity.
We further assume that the predictive model essentially estimates this order parameter $y(p)$ to reconstruct the physical system's parameters, i.e. make the prediction $p_{\mathrm{pred}}$. As illustrated in Fig.~\ref{fig:method}(c), let us consider three neighboring points $x_1$, $x_2$ and $x_3$ on an equidistant grid in one parameter dimension. 
As the sampling in parameter space is uniform, the probability $W$ of each of these points is the same, i.e. $W(p_0=x_1)=W(p_0=x_2)=W(p_0=x_3)$. If the resolution of the model is on the order of the distance between these points, we neglect the probabilities of the model to err by more than one point on the given grid so that $\sum_{n=1}^3 W(p_{\mathrm{pred}}=x_n | p_0=x_2)=1$
 for the probability conditioned on $p_0=x_2$.

The probabilistic sampling of the finite dimensional physical system gives rise to randomness, furthermore the predictive model will typically have random errors.
Therefore, the estimation $y_E(S)$ of the order parameter as inferred by the model for different samples at a parameter $p_0$ will generally vary probabilistically around the expectation value $\langle y(p_0) \rangle$. 
For a simplified description, we will assume the deviation $y_E - \langle y(p_0) \rangle$ to be symmetric around zero with lower probability density for higher deviations and to
follow independent identical distributions at all points
\footnote{The first condition corresponds to the reasonable assumption of an unbiased estimator. The second condition is necessary to establish a concrete analytical model but could be altered to develop a more general description.}.

Under these conditions the
 predictive model will bias its predictions in the direction where the order parameter varies less
\begin{align}
&{W(p_{\mathrm{pred}}=x_1 | p_0=x_2) }>{W(p_{\mathrm{pred}}=x_3 | p_0=x_2) } \nonumber \\
&\Leftrightarrow 
|y(x_1)-y(x_2)| < |y(x_3)-y(x_2)|.
\label{eq:condition}
\end{align}
Obviously, this leads to diverging $\delta p$ at a first-order phase transition, where $y$  jumps discontinuously. Taking into account a finite system size or considering second-order phase transitions, the direction of $\delta p$ will depend on the sign of the second derivative of $y$, which will generally change when going from one phase to another. Depending on the particular transition, this sign can also change a second time, leading to another peak in the divergence. Therefore further peaks in divergence can be expected and each peak is indicator, but not a proof of a phase transition.

These arguments show the potential of our method in idealized limiting cases of high and low precision.  To evaluate how this translates to a  non-ideal
 setting and thereby test the performance of our scheme, we apply it to two physical systems, which are introduced in the next section.

\section{Test cases}

\begin{figure}[t!]
\includegraphics[width=0.47\textwidth]{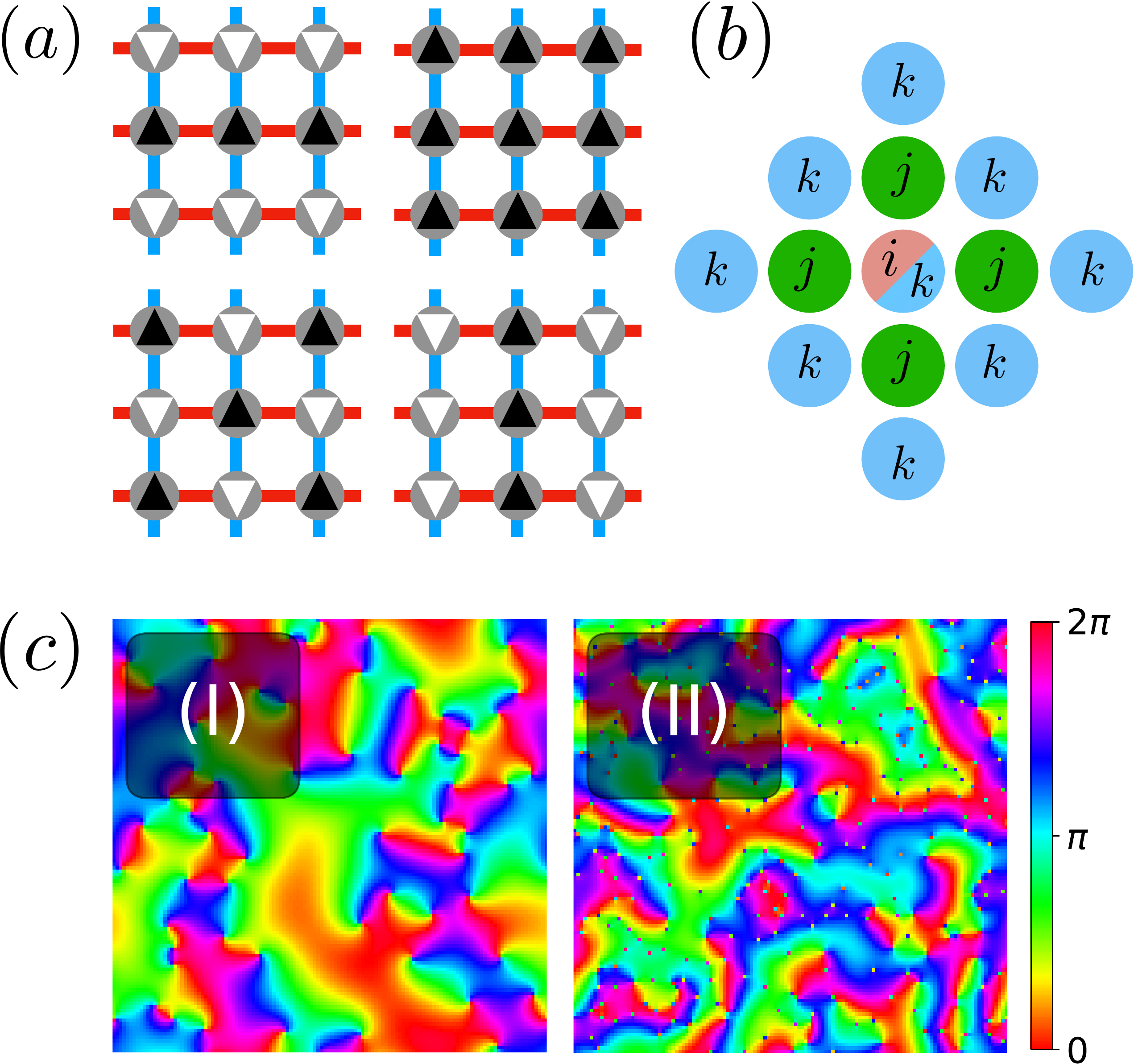}
\caption{(a) Ising model on a two-dimensional lattice with coupling $J_x$ in horizontal direction (red) and coupling $J_y$ in vertical direction (blue). For large enough coupling strength as compared to the temperature, the system will enter an ordered phase, whose spin configuration depends on the sign of couplings. The four different ordered phases are displayed with $J_x>0$ on the right, $J_x<0$ on the left, $J_y>0$ on top, and $J_y<0$ at the bottom, cf. Eq. \eqref{eq:IsingH}.
(b) Kuramoto-Hopf model with nearest neighbor (green) and next-to-nearest neighbor (blue) couplings of the $i$-th oscillator (red) on a two dimensional array. While $V_1$ and $C$ couple only nearest neighbors, the parameter $V_2$ includes coupling to the next-to-nearest neighbors. 
(c) The patterns (I) and (II) are samples of stationary phases for different values of $V_2$  on a $128\times128$ array with random initial conditions for constant $C=1$ and $V_1=5$, cf. Eq. \eqref{eq:hopf}. Panel (I) shows the spiral structures for the case $V_2/C = 0.1$. For $V_2/C \approx 4$ several stable $\pi$-defects emerge in panel (II).
}
\label{fig:systems}
\end{figure}

{

\subsection{The Ising model}

As a basic test case to demonstrate our method, we apply it to the well-understood  Ising model \cite{ising1925}.
Perhaps its most popular version is the case of
 ferromagnetically coupled spins on a two-dimensional lattice without bias fields and isotropic coupling. It is described by the Hamiltonian 
\begin{equation}
H_{\mathrm{Ising}}^\mathrm{iso}= -J \sum_{j=1}^N \sum_{k=1}^N  \left( \sigma_{j,k} \sigma_{j, k+1} + \sigma_{j+1,k} \sigma_{j, k} \right),
\label{eq:IsingISO}
\end{equation}
where the $\sigma_{jk}$ describe spin configurations that can take on the values $\sigma_{jk}=\pm 1$ and $J>0$ denotes the coupling energy between neighboring spins.  For temperatures above the critical value, 
the magnetization $M=\sum_{j,k} \sigma_{jk} /N^2$ vanishes in the thermodynamic limit $N \to \infty$. 
At $T_c$ the system undergoes a phase transition, with a net magnetization building up, that reaches $|M|=1$ at $T=0$.

We will also consider the natural generalization
described by the Hamiltonian
\begin{equation}
\label{eq:IsingH}
H_{\mathrm{Ising}}^\mathrm{aniso}= -\sum_{j=1}^N \sum_{k=1}^N  \left( J_x \sigma_{j,k} \sigma_{j, k+1} + J_y \sigma_{j+1,k} \sigma_{j, k}\right),
\end{equation}
allowing for different coupling strength $J_x$ and $J_y$ in horizontal and vertical direction, as well as negative values for the coupling. 
In this anti-ferromagnetic case, it is energetically better to anti-align neighboring spins.
The threshold for the absolute value of the coupling strength above which the ordered phase emerges, now depends on both $J_x$ and $J_y$ 
 and was derived analytically \cite{onsager44} to be $J_y/T=-\log(\tanh(J_x/T))/2 $ for $J_x, J_y >0$, where we set the Boltzmann constant $k_B=1$.
The other three sectors, where the sign of at least one of the couplings is negative, behave analogously with adjusted order parameter
$M=\sum_{j,k} \sigma_{jk} \mathrm{sign}(J_x) \mathrm{sign}(J_y) /N^2$, measuring anti-alignment instead of alignment of spins, where appropriate. The different couplings and the corresponding ideal configurations at $T=0$ are illustrated in Fig.~\ref{fig:systems}(a).

Numerically, we generate samples of the Ising model on a $512\times512$ lattice using the Metropolis algorithm \cite{metropolis_1953}:
 The lattice is initialized in a random configuration and then updated many times by drawing a random spin, which is then flipped with probability
 $\min \left(1,\mathrm{e}^{-\Delta E / T}\right)$, where $\Delta E$ is the energy difference resulting  from the considered flip. To ensure that the system is sufficiently thermalized we sweep the complete lattice $10^5$ times, where each point is updated once per sweep.

}

\subsection{The Kuramoto-Hopf model}

To try our method on a new, relatively unexplored system, we choose the Kuramoto-Hopf model \cite{marquardt2015} which describes the dynamics of weakly-coupled limit-cycle oscillators on a square lattice. It could be experimentally implemented e.g. with optomechanical self-oscillators, for which it was first derived \cite{marquardt2011}.
In the situation of weak coupling and linear nearest-neighbor interactions, the Hopf equations for limit cycle oscillators with amplitude-dependent frequency give rise to the
effective equations
\begin{align}
\dot{\varphi}_i &= C \sum_{\langle j,i\rangle}\cos (\varphi_j-\varphi_i)+V_1 \sum_{\langle j,i\rangle}\sin (\varphi_j-\varphi_i)\notag\\
&+V_2\left\{\sum_{\langle j,i\rangle}\sum_{\langle k,j\rangle} \left[\sin(2\varphi_j-\varphi_k-\varphi_i)-\sin(\varphi_k-\varphi_i)\right]\right.\notag\\
&+\left.\sum_{\langle j,i\rangle}\sum_{\langle k,i\rangle}\sin(\varphi_k+\varphi_j-2\varphi_i)\right\},
\label{eq:hopf}
\end{align}
for the phases $\varphi_i$ of the oscillators. 
Equation \eqref{eq:hopf} describes both nearest-neighbor couplings $\propto C, V_1$ and a more complex type of coupling including both nearest-neighbor and next-to-nearest couplings $\propto V_2$.
The structure of the coupling is illustrated in Fig.~\ref{fig:systems}(b), where the indices appearing in Eq. \eqref{eq:hopf} are colored according to their roles.

 The interplay of the different terms leads
to rich phase patterns.
For example, at fixed $V_2=0$, increasing the ratio $V_1/C$ interpolates from the Kuramoto-Sakaguchi model \cite{sakaguchi_1986} to the Kuramoto model \cite{kuramoto_1975,acebron_2005} at $V_1 \gg C$. 
In the present article, we focus on varying the next-to-nearest neighbor coupling parameter $V_2/C$ at constant $V_1/C=5$. Thereby, we interpolate between a phase that has a well-defined continuum limit at  $V_2=0$, and a phase where the next-nearest-neighbor terms become important. There, $\pi$-defect configurations, where the phase of a single oscillator within the lattice is opposite to its direct neighbors become stable \cite{marquardt2015}. 
Examples of the phase patterns in the different phases are
shown in Fig.~\ref{fig:systems}(c).

Samples of the Kuramoto-Hopf model are obtained by propagating Eq. \eqref{eq:hopf} until time $t=500/C$ with random initial states on a $128\times128$ grid using the DifferentialEquations.jl package of julia
 \cite{julia, juliaDE} with the CVode Backward Differentiation Formula (BDF). Within the Newton iteration for this implicit method, we use GMRES as the linear solver  \cite{juliasundials}.

\begin{figure}[h!]
\includegraphics[width=0.49\textwidth]{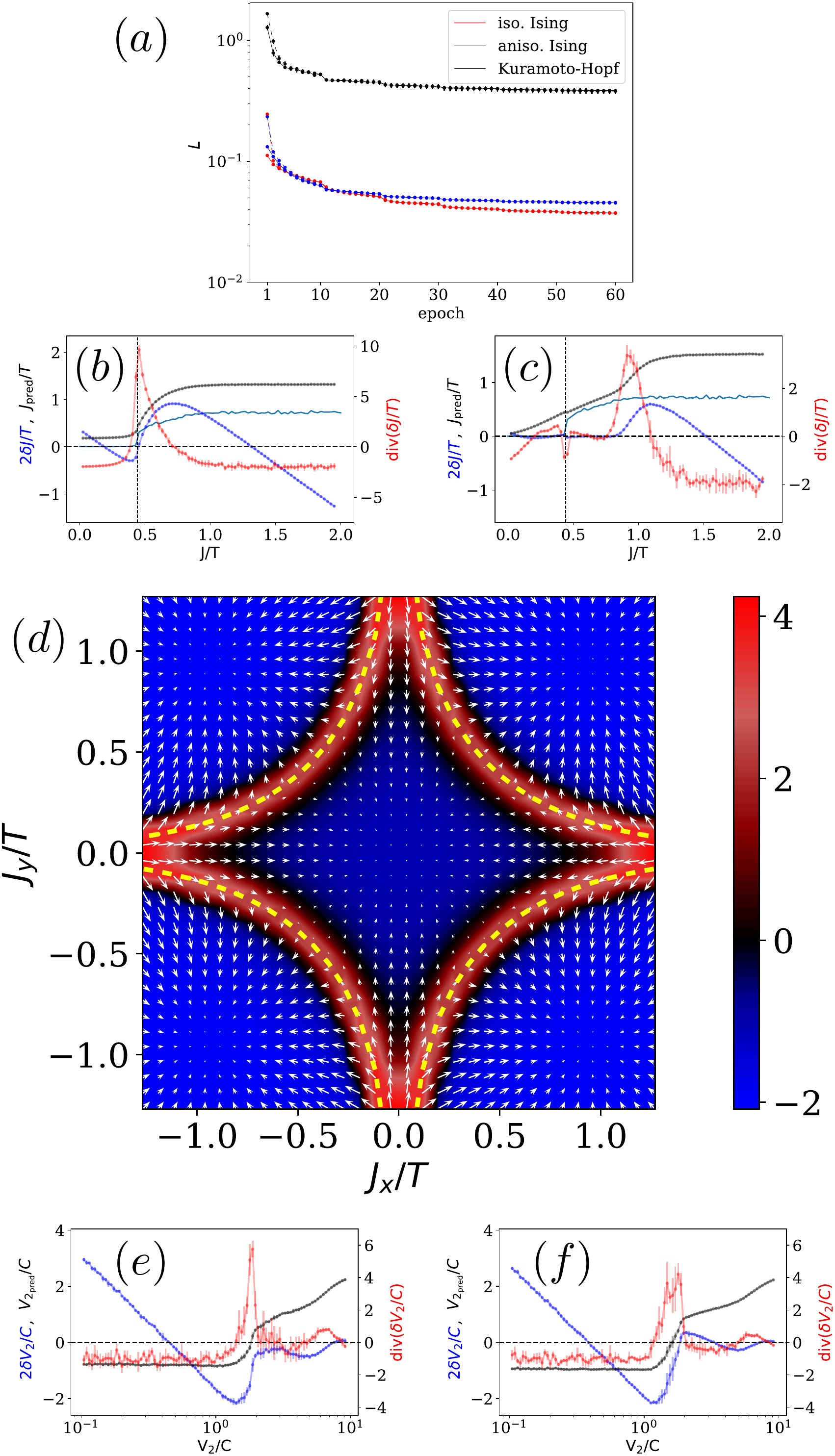}
\caption{ Panel (a) is a plot of the loss function $L$ from Eq.~\eqref{eq:loss} as a function of the number of training epochs for both Ising and Kuramoto-Hopf model. The resulting deviation of the predictions (blue) and their divergence (red) from  \eqref{eq:div} for the coupling constant $J$ of the isotropic Ising model are plotted in panel (b) after one learning epoch and panel (c) after 60 epochs. The dashed black line indicates the analytical value $J/T=0.44$ of the phase transition. The error bars are obtained by splitting a total of 442 samples per point into 7 sets. 
For the anisotropic Ising model, panel (d) shows the scaled deviations $\delta J$ (white arrows) after the first training epoch, and the resulting divergence as a function of $J_x$ and $J_y$. The dashed yellow lines again indicate the Onsager result of the phase transition. Here, a total of 100 samples per point were split into 10 sets.
Panels (e) and (f) show the network's predictions (black), the resulting deviation of the predictions (blue) and their divergence (red) from \eqref{eq:div} for the coupling constant $V_2/C$ of the Kuramoto-Hopf model after five learning epochs and after 60 epochs, respectively. For this case we used 66 samples per point that were split into 3 sets.
For all models, each set was trained 50 times from scratch for further averaging.
}
\label{fig:results}
\end{figure}

\section{learning scheme applied to the systems}

While our method does not require a particular predictive model, in this work we use neural networks.  Their basic architecture is of the form $m= \Pi_{n=1}^L  K_n A_n$ for an $L$-layer network,
with linear functions $A_n$ and nonlinear functions $K_n$ of the features. As the couplings of both considered models act only locally, cf. Fig.~\ref{fig:systems},
the resulting samples have a local structure just like ordinary images. Therefore we use convolutional neural networks, where the first few $A_n$ perform convolutions, and the later layers have all-to-all connectivity. The $K_n$ are rectifiers, i.e. $K_n$ applied to a vector $v$ changes the $j$th entry according to $v^j \to \max(0, v^j+b^j_n)$, where $b$ is a bias term learned by the model to fit the data. For this fit,
we use backpropagation with the Adam optimizer \cite{adam} to train the model in a series of epochs, where the network sees each training image once per epoch.  The learning rate of the optimizer is halved every five epochs. Our particular network consists  of three convolutional layers followed by three fully-connected layers, with rectifiers activation functions in the hidden layers.

The samples for each data point are split in two sets of equal size: a training set for training the network and a test set, on which it is evaluated and the predictions are calculated. To obtain error bars, we repeated this process on several data sets. 
The average are collected in Fig.~\ref{fig:results}: To give an overview of the learning procedure, panel (a) shows the loss function $L$ from Eq. \eqref{eq:loss} for both Ising models and the Hopf-Kuramoto model as a function of passed epochs.

Panel (b) shows the network's output after two epochs for the isotropic Ising model from Eq.~\eqref{eq:IsingISO}. Samples as a function of $J/T$ are obtained on a linearly spaced grid of 80 points on the interval $[0, 2]$, where each point was sampled 440 times.
A clear peak of the divergence (Eq.~\eqref{eq:div}) emerges close to the analytically expected value for the Ising phase transition at $J/T\approx 0.44$. A closer inspection of the predictions $J_{\mathrm{pred}}(J)$ suggest that at this epoch, the network's behavior is well described by the situation explained in Fig.~\ref{fig:method}(b), i.e. the network
 essentially recognizes whether the sample is in the ferromagnetic or in the paramagnetic phase, but cannot resolve well within the phases. 
Panel (c) shows the predictions after the final training epoch, where the network has learned to resolve the parameters even within a phase in the region $J <T_c$. 
The higher resolution strongly decreases the size of the peak, but the maximum is still in the vicinity of the phase transition, as we would expect from the arguments illustrated by Fig.~\ref{fig:method}(c).
Furthermore, a second peak arises at around $J=T$, which is approximately where the magnetization (shown in the same plot) starts to saturate. After saturation the states again become indistinguishable to the model.

Panel (d) shows the output of the network after two training epochs for the  Ising model from Eq.~\eqref{eq:IsingH} with independent couplings $(J_x, J_y)$ sampled on a $40\times40$ grid with each parameter ranging from $-1.5$ to $+1.5$. The divergence peak is in good agreement with the Onsager result given below  Eq.~\eqref{eq:IsingH}, which is represented by the yellow lines. The arrows show scaled deviations $\delta \vec J$ to indicate their direction. The predictions themselves are again clustered at the center of the different phases, similar to the situation expected from Fig.~\ref{fig:method}(b).

For the Kuramoto-Hopf model, we sampled a line in the phase diagram at $V_1/C=5$ and $V_2/C$ logarithmically varying between $V_2/C=0.1$ and $V_2/C=10$. We discretized the line by 100 grid points and we computed 70 independent samples per grid point.  
The $\pi$-defects mentioned earlier become stable at around $V_2/C \approx 1$ \cite{marquardt2015} manifesting a phase transition. 
Panels (e) and (f) show the output of the network after, respectively, 5 and 60 training epochs. At 5 epochs, there is a clear peak centered at approximately $V_2/C \approx 2$, after 60 epochs it becomes broader and starts at slightly lower values. While the phase transition is captured, the exact value of the transition is less precise than in case of the Ising model. Inspecting the samples around $V_2/C \gtrsim 1$, we find that only a few of them actually show defects due to the finite size of the lattice.

\section{Discussion and Outlook}

For both of the systems tested, the learning scheme showed promising results
and found the respective phase transitions without any prior knowledge. During the training procedure, as the resolution of the neural network increases, more and more structure in the phase diagram emerges. It is important to avoid over-fitting to the training data, as this may lead to spurious structures emerging in $\delta p$ related to the features of the particular training data. Obviously over-fitting can be avoided by obtaining more training data, but also by data augmentation, where suitable random transformations are applied each time a sample is revisited. In our case, we made use of the symmetries of the systems and applied a series of flips, rotations by multiples of 90 degrees, as well as random translations along both the $x$ and $y$ axis with periodic boundary conditions.
Finally, the training should be stopped early enough, as inferred from comparing the difference between validation loss and training loss. 

To achieve a higher resolution of the phase boundaries, it is helpful to have more training data. In practice the amount of data is limited by the availability of computational resources to generate them. Assuming therefore a fixed total number of samples, the number of samples per grid point scales as $1/\Delta^M$ with distance $\Delta$ between adjacent grid points of sampled system parameters on a $M$-dimensional grid. As the divergence is numerically obtained via a difference quotient with $\Delta$ in the denominator, the divergence's confidence interval scales in total as $1/\Delta^{M/2+1}$. Furthermore, the divergence scales with the magnitude of $\delta p$, so that a higher resolution of the predictive model reduces the signal. As the noise may not decrease accordingly and may even be dominated by issues such as the inherent stochasticity of the training procedure, a higher resolution for $p_{\mathrm{pred}}$ does not necessarily lead to a better resolution of the phase transition. 

These considerations should be kept in mind when the method introduced here will be applied to different systems. Several possibilities arise for further development of the method in the future. For example, to address the trade-off in resolution of $p_{\mathrm{pred}}$ and the actual phase transition, one could add a term to the loss function which diverges at $p_{\mathrm{pred}}=0$ in order to prevent the predictive model from decreasing $p_{\mathrm{pred}}$ too much. In the limit where the network's resolution is already very high, it may be advantageous to view the different grid points of the physical parameters as categories to be learned rather than parameters to be continuously estimated. The optimal configuration with respect to these changes may differ between different systems.
It is also worth  exploring other functions (besides the divergence) to quantify how the deviations $\delta \vec p$ point in different directions near phase boundaries. 
Finally, one could train a neural network to infer the system parameters of samples, but ignore the output layer, and rather use the last hidden layers as a representation of the data. This representation can then be used  as an input for clustering algorithms.

As our scheme is easy to use and allows for the assignment of phase labels without any prior knowledge, it can be used as a starting points for other methods that do require some initial knowledge of the phase diagram's structure, such as \cite{huembeli2018}.
Furthermore, once the phase labels are determined, algorithms that visually interpret the decision making of neural networks \cite{zhang2018visual} can be used to try and retrieve the relevant features distinguishing the different phases.

\section{Conclusion}

In summary, we have introduced the divergence-based learning scheme of phase transition and provided a theoretical description of the method as a guide to its application, which was successfully demonstrated on both an equilibrium system, the Ising model, and a non-equilibrium system, the Kuramoto-Hopf-model.
As our method can be built upon any predictive model and only requires the calculation of the divergence, it is simple to use and implement with existing machine learning libraries.
Because that predictive model needs to be trained only once on the training set, the method is economical in computational resources.
It is applicable in the generic situation where no prior information of the phases is known and for phase diagrams of arbitrary parameter dimension.
Due to these properties, we hope it will become a useful part of physicists' toolbox to discover novel phase transitions.

\section{Acknowledgments}
We would like to thank C. Bruder and J. Lehmann for helpful discussions. 
This work was financially supported by the Swiss National Science
Foundation (SNSF) and the NCCR Quantum Science and Technology. Calculations were performed at sciCORE (scicore.unibas.ch) scientific computing core facility at University of Basel.

\bibliography{ml_bibliography}

\end{document}